\documentclass[showpacs,prb,twocolumn]{revtex4}
\usepackage{textcomp}
\usepackage{amsmath}
\usepackage{amssymb}
\usepackage{graphicx} 
\usepackage{epsfig}
\usepackage[english]{babel}
\usepackage{amsfonts}

\newcommand{\beq}{\begin{equation}}
\newcommand{\eeq}{\end{equation}}
\newcommand{\Ham}{\mathcal H}

\begin{document}

\title{Spin-supersolid phase in Heisenberg chains: a characterization via Matrix Product States 
with periodic boundary conditions}

\author{Davide Rossini}
\affiliation{Scuola Normale Superiore, NEST and Istituto Nanoscienze - CNR, Pisa, Italy} 

\author{Vittorio Giovannetti}
\affiliation{Scuola Normale Superiore, NEST and Istituto Nanoscienze - CNR, Pisa, Italy} 

\author{Rosario Fazio}
\affiliation{Scuola Normale Superiore, NEST and Istituto Nanoscienze - CNR, Pisa, Italy} 

\begin{abstract}
  By means of a variational calculation using Matrix Product States with periodic boundary 
  conditions, we accurately determine the extension of the spin-supersolid phase predicted 
  to exist in the spin-1 anisotropic Heisenberg chain. We compute both the structure factor 
  and the superfluid stiffness, and extract the critical exponents of the supersolid-to-solid 
  phase transition.
\end{abstract}



\maketitle

A phase of matter where diagonal (solid) and off-diagonal (superfluid) long-range order coexist is 
named supersolid. Since its original prediction,~\cite{originalsupersolid}
the search for this phase has attracted the attention of a growing number of experimental and theoretical 
physicists.~\cite{prokofiev07} However, despite this great effort, the supersolid phase has, to date, 
eluded a firm experimental confirmation. This is due to the fact that the stabilization of such a phase arises 
from the combined action of two mutually exclusive effects: on one side, the solid order 
requires a well defined spatial arrangement of the atoms in real space; on the other side, 
the superfluid order requires the atoms to be delocalized and condensed in a macroscopic quantum state. 

The first, and probably most prominent, candidate for the experimental 
realization of a supersolid phase is~$^4$He.~\cite{kim04}
More recently, the trapping of cold atoms in optical lattices has stimulated the search
for such exotic phase in these systems (see Refs.~\onlinecite{coldatoms} and references therein).

Furthermore, in strict analogy with what postulated in the fields of quantum fluids and cold atomic gases, 
a {\em spin-supersolid} phase can be defined also in the context of quantum magnets, in association with 
a simultaneous ordering along the z-direction at finite momentum and of a breaking of U(1) symmetry in the xy-plane. 
Examples of such phases have been found~\cite{ng06,laflorencie07,picon08} in $S = 1/2$ spin-dimer model on a square 
lattice, where extra singlets delocalize in a solid background via correlated hoppings,~\cite{picon08} 
and in $S=1$ systems.~\cite{sengupta07a,sengupta07b,peters}

The spin-$1$ Heisenberg chain with a single-site uniaxial anisotropy in a transverse magnetic 
field is what we study in the present paper. For this model, Sengupta and Batista~\cite{sengupta07b} 
predicted a spin-supersolid phase for intermediate values of the external field and of the uniaxial anisotropy. Their 
analysis of the phase diagram was based on the derivation of an effective model and on Quantum Monte Carlo 
simulations. Further confirmation using Density Matrix Renormalization Group (DMRG) was 
reported in Ref.~\onlinecite{peters}. In these last works, the existence of the supersolid phase 
was inferred by an analysis of the magnetization profiles. However, due to the intrinsic limitation 
of standard DMRG techniques to the case of Open Boundary Conditions (OBC), it was impossible to access 
the superfluid order parameter with such kind of algorithm.
A detailed numerical analysis of the supersolid phase would indeed require the simultaneous study 
of both diagonal and off-diagonal orderings. The Matrix Product States (MPS) approach to DMRG,~\cite{verstraete08} 
with its recent generalization to study efficiently one-dimensional systems with Periodic Boundary 
Conditions (PBC),~\cite{verstraete04,sandvik07,pippan10,pirvu10,rossini11} 
appears to be an ideal tool to determine such parameters.
Here we exploit this fact to address both the diagonal and off-diagonal order parameters
for the spin-$1$ Heisenberg model of Refs.~\onlinecite{sengupta07b,peters}:  
this allows us to directly access the so called {\em spin stiffness} of the system, 
and therefore to accurately locate the supersolid phase. 

The model under investigation is governed by the following spin-1 Heisenberg Hamiltonian
\begin{eqnarray}
  \Ham &=& \sum_j \left[ \frac{J}{2} \left( S^+_j S^-_{j+1} + {\it h.c.} \right) + \Delta S^z_j S^z_{j+1} \right] \nonumber\\
  &+& D \sum_j (S^z_j)^2 - B \sum_j S^z_j \,,
  \label{eq:Spin1}
\end{eqnarray}
where $S^\alpha_{j}$ (with $\alpha=x,y,z$) are the spin-$1$ operators for the $j$-th site, 
while $S^\pm_j$ are the associated raising/lowering operators;
PBC are imposed by requiring $S^\alpha_{N+1} = S^\alpha_1$ ($N$ is the number of sites in the chain).
Notice that, in addition to the exchange coupling $J$ and the magnetic anisotropy $\Delta$, 
the model also includes a coupling to an external transverse field $B$ and a single-site 
uniaxial anisotropy of strength $D$. Hereafter we set $J=1$, thus fixing the energy scale. 
Furthermore, following Ref.~\onlinecite{sengupta07b},
we set $\Delta = 2D$. Units of $\hbar = k_b = 1$ are used.

\begin{figure}
  \begin{center}
    \includegraphics[width=\columnwidth]{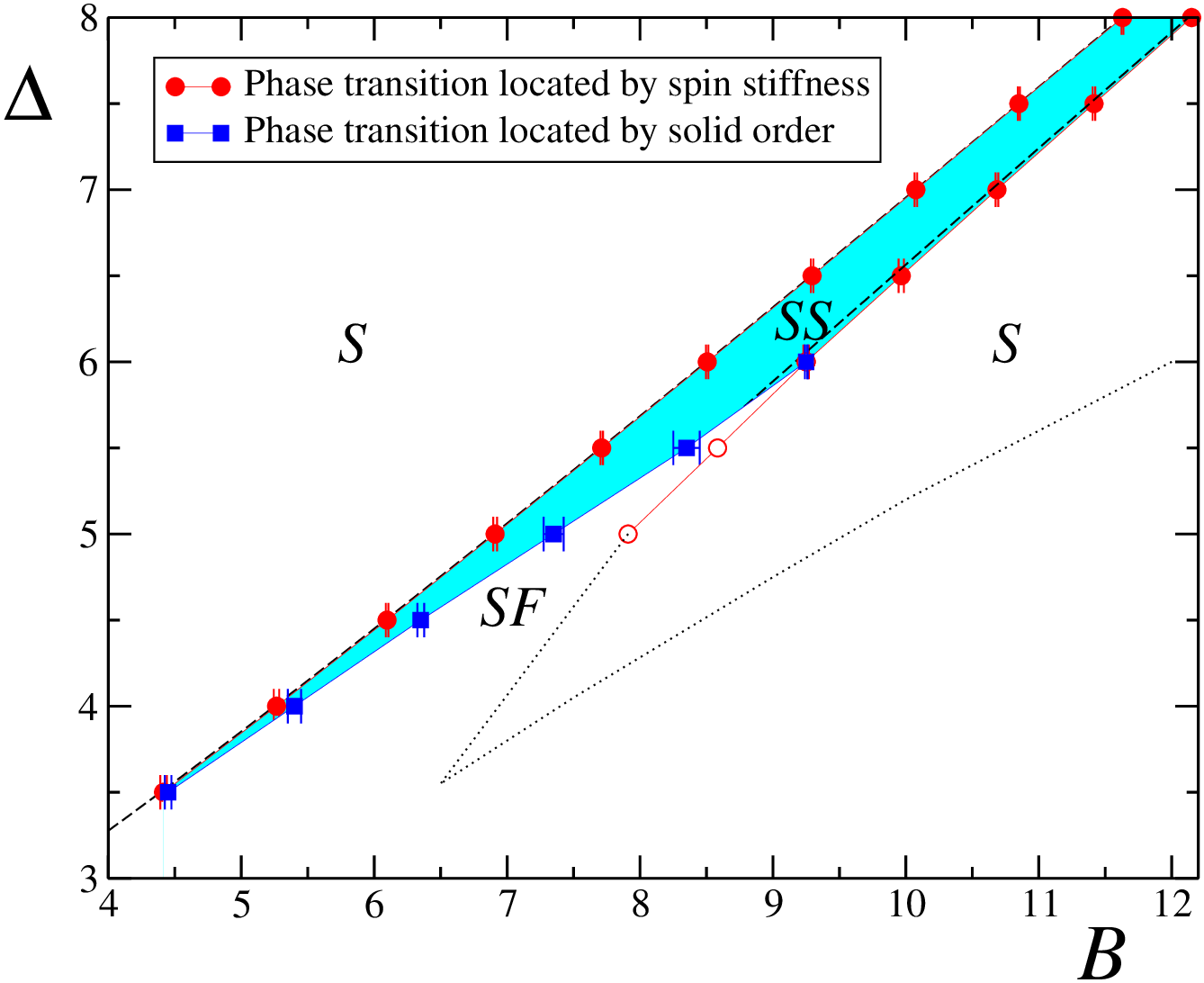}
    \caption{(color online). Extension of the supersolid phase (cyan region) in the $\Delta - B$ plane, 
      for a one-dimensional anisotropic spin-$1$ Heisenberg Hamiltonian  
      in a transverse magnetic field, and single-site uniaxial anisotropy defined in Eq.~\eqref{eq:Spin1}.
      The value of uniaxial anisotropy is fixed at $D = \Delta/2$.
      The phase boundaries are located by evaluating the region of parameters for which 
      the solid order parameter ${\mathcal O}_{SDW}$ and the spin-stiffness $\rho_s$ of Eq.~\eqref{eq:stiff}
      were simultaneously different from zero. For $\Delta \gtrsim 6$, where the transition is between 
      the solid and the supersolid, the vanishing of the superfluid stiffness is an excellent indicator 
      of the supersolid boundaries. For smaller values of $\Delta$ the transition is to a superfluid phase, 
      therefore here the boundary of the supersolid is determined by the vanishing of the 
      solid order (blue squares), while the spin stiffness vanishes at larger values of the external 
      field $B$ at the boundary between the superfluid and the spin-gapped phase 
      (open circles). The two dashed lines are the result of effective low-energy models, valid for 
      $\Delta \gg 1$.~\cite{sengupta07b} The dotted lines are directly taken from the simulations 
      of Ref.~\onlinecite{sengupta07b} and separate the solid phase from the superfluid region 
      at large values of $B$ and small values of $\Delta$.
      In the figure $S$ = Solid, $SS$ = Supersolid, $SF$ = Superfluid.}
    \label{fig:spin1_PhDiag}
  \end{center}
\end{figure}

The phase diagram described by the model in Eq.~\eqref{eq:Spin1} is quite rich (see, e.g., Fig.~\ref{fig:spin1_PhDiag}). 
For large values of the anisotropy $\Delta$, it goes into a spin-gapped Ising-like phase showing 
long-range diagonal order. On increasing the external field, there is a transition to a superfluid phase 
characterized by a finite spin-stiffness. At larger values of $B$, the system goes into a fully polarized 
state (not shown in Fig.~\ref{fig:spin1_PhDiag}). In between the spin-gapped and the superfluid 
phase, a spin-supersolid was shown to exist,~\cite{sengupta07b} possessing simultaneously diagonal 
and off-diagonal ordering. We concentrate on this specific configuration.

The solid ordering can be detected by an analysis of the spin-structure factor, defined as 
\beq
   S^{zz}(q) = \frac{1}{N} \sum_{j,\ell} e^{-i q (j-\ell)} \langle S^z_j S^z_\ell \rangle \;,
   \label{eq:StructureZZ}
\eeq
at momentum $q = \pi$. A solid order parameter can be defined as 
${\mathcal O}_{SDW} = \lim_{N \to \infty} \frac{S^{zz}(\pi)}{N}$: indeed non zero values of this quantity 
indicate that the dominant correlations have a Spin Density Wave (SDW) character.
Off-diagonal order instead is detected by the superfluid stiffness, defined as
\beq
   \rho_s = N \frac{\partial^2 E_0(\phi)}{\partial \phi^2} \Big\vert_{\phi = 0} \, ,
   \label{eq:stiff}
\eeq
where $E_0(\phi)$ is the ground state energy of the chain with twisted boundary conditions, 
or equivalently~\cite{fisher73} in the case where $J \longrightarrow J e^{i \phi/N}$. 
For PBC $\rho_s$ quantifies the system's response to an infinitesimal
magnetic flux which is added through the ring.
Vice-versa for OBC it nullifies, since the twist $\phi$ 
can be wiped out by a gauge transformation.
The simultaneous nonzero values of Eqs.~\eqref{eq:StructureZZ} and~\eqref{eq:stiff} 
signal the supersolid phase. 
Our investigation leads to the result summarized in Fig.~\ref{fig:spin1_PhDiag}. 
In the following we provide detailed evidence of this result.

\begin{figure}
  \begin{center}
    \includegraphics[width=\columnwidth]{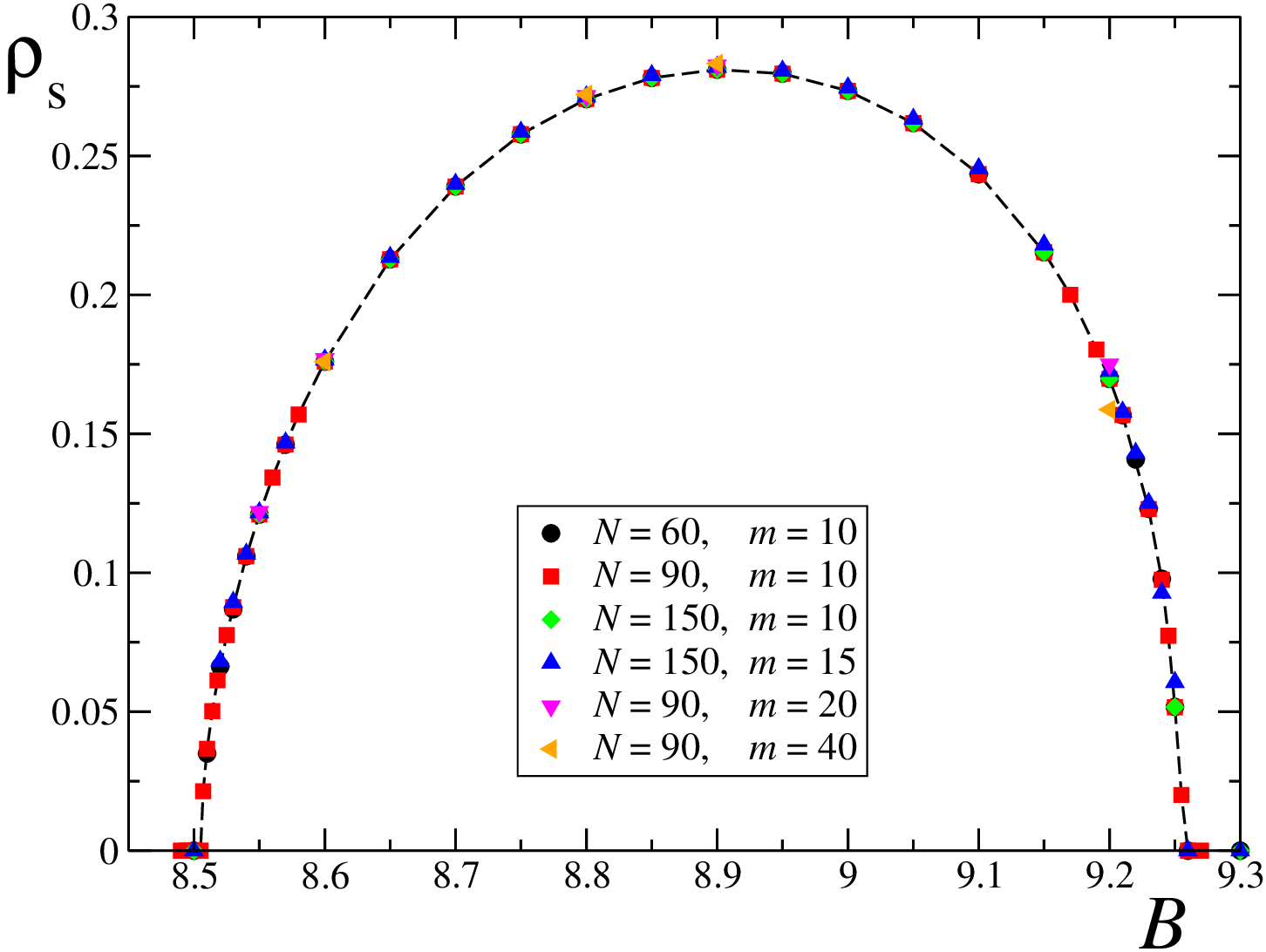}
    \caption{(color online). Spin stiffness for model~\eqref{eq:Spin1} with $\Delta = 6$ and $D = \Delta/2 = 3$,
      in a parameter range where the system is in a supersolid phase.
      Parameters used in the MPS variational wavefunction for the various sets of data are as follows:
      for $m = 10$ we performed $n_s = 30$ sweeps, with truncation parameter $p = 25, \, 20, \, 15$, 
      respectively for $N = 60, \, 90, \, 150$;
      for $m = 15, \, N = 150$ we used $p = 30, n_s = 40$;
      for $m = 20, 40$ we respectively took $p=45,60$ with $n_s=50$.
      We kept $s=40$ in all the cases except for $m=40$, where $s=60$, 
      obtaining comparable precisions 
      in the energy fluctuations for each of those parameter settings.}
    \label{fig:Stiff_Spin1}
  \end{center}
\end{figure}

Our algorithm is based on Refs~\onlinecite{pippan10,rossini11}, where details of the implementation can be found. 
We considered chains up to $N=180$, where no finite-size effects could be detectable for our precisions. 
The dimension of the matrices used in the MPS ansatz with PBC was taken up to $m=40$, while the minimization
of the ground state energy was obtained by optimizing the structure site by site,
sweeping through the ring in a circular fashion with a sufficient amount $n_s$ of sweeps.
As discussed by Pippan {\em et al.},~\cite{pippan10} an important speedup in the code can be achieved by 
introducing a factorization procedure for long products of MPS matrices, which reduces the computational effort. 
Intuitively this is justified by the fact that, for large chains, the local physics
of the system is not affected by the properties of the boundaries.
The degree of this factorization is characterized by {\em two} truncation parameters $p$ and $s$,~\cite{note1} 
that in our simulations were taken to be $10 \lesssim p,s \lesssim 60$ (for a formal 
definition of these quantities we refer the reader to Ref.~\onlinecite{rossini11}). 
We checked that our choice of $m$ and $p$ would guarantee the convergence of our results. 

For the calculation of the stiffness, we computed the dependence of the ground state energy 
as a function of the twist and then fitted the curve  with a quadratic law 
$E_0(\phi) = E_0(0) + c_2 \phi^2$, obtaining the prefactor $c_2$ which is directly related 
to the stiffness: $\rho_s = 2 N c_2$. The determination of the boundary for the solid order 
turned out to be more demanding, due to the necessity to measure long-range spin correlations, 
i.e., the quantities $\langle S^z_j S^z_{j+r} \rangle$ of Eq.~\eqref{eq:StructureZZ}, for $r \gg 1$.
Contrary to the evaluation of ground-state energies that enter in Eq.~\eqref{eq:stiff}, 
this generally requires a high degree of accuracy in the MPS representation of the ground state, 
thus implying large values of $m$. To enhance the precision, we hence used the fact that the solid order
in the bulk of the system is not qualitatively affected by the choice of OBC or PBC, and 
ran simulations using MPS with OBC~\cite{verstraete08} which allows one to  work with 
matrices of larger dimension (i.e., with $m$ of order $100$). We also carefully checked 
that the obtained results were not plagued by finite-size corrections.

\begin{figure}
  \begin{center}
    \includegraphics[width=\columnwidth]{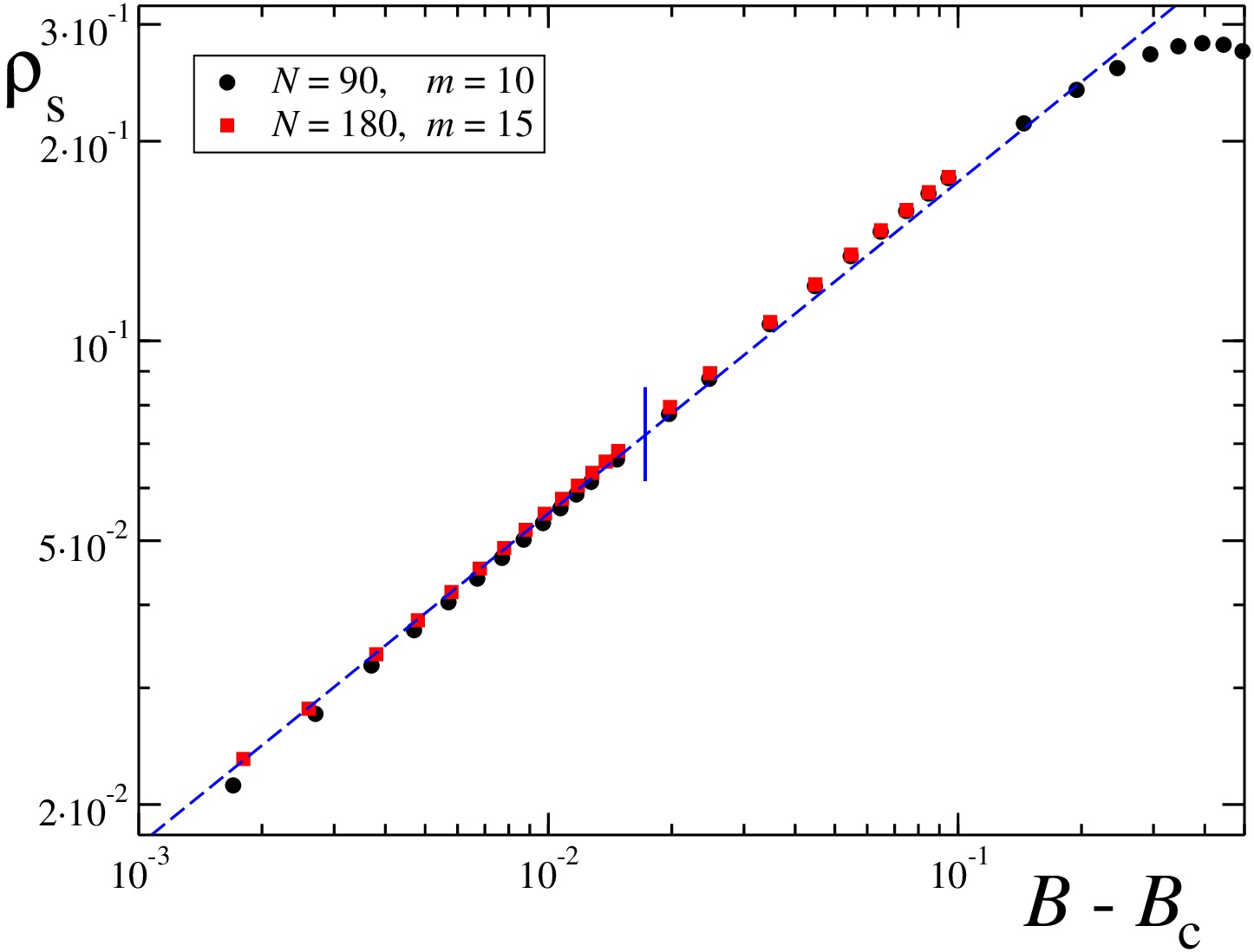}
    \caption{(color online). Scaling of the spin stiffness for the supersolid-to-solid transition at $\Delta = 6$. 
      The critical point is estimated to be $B_c = 8.5052 \pm 0.0005$. 
      Black circles is the same data set for $m=10, \, N = 90$ in Fig.~\ref{fig:Stiff_Spin1}.
      Red squares are for $m=15, \, N = 180$, with $p = 30$, $s=40$, and $n_s = 50$.
      The scaling is compatible with a power-law behavior of exponent $\beta_s = 0.5$
      (dashed blue line, plotted as a guidance).
      The power-law fits of the two data sets until the vertical blue line,
      respectively giving $\beta_s = 0.521$ and $0.511$,  confirm this prediction.}
    \label{fig:Stiff-scaling}
  \end{center}
\end{figure}

Our findings are summarized in Fig.~\ref{fig:spin1_PhDiag}, which details the phase diagram of the system 
obtained by computing the solid and superfluid parameters ${\mathcal O}_{SDW}$ and $\rho_s$ for different values 
of $B$ and $\Delta$. Consider first the results we obtained for the superfluid stiffness focusing 
on a single value of the anisotropy, say $\Delta=6$.
The behavior of $\rho_s$ for such value of $\Delta$ is summarized in Fig.~\ref{fig:Stiff_Spin1},
where a cusp-like shape for $\rho_s$ as a function of $B$ emerges: 
in the critical region between $8.51 \pm 0.01 \lesssim B \lesssim 9.25 \pm 0.01$ the superfluid phase 
is present, as testified by the fact that here $\rho_s$ is not null. 
For most values of the magnetic field, modest values of $m$ seem to be sufficient to attain 
good accuracies; close to the border of the critical zone, 
where variations of $\rho_s$ are more sensitive upon increasing $m$,
higher precision are required though.
For all the considered values of $m$, the errors are smaller than the symbol size.
As an example, for $B=8.6$, ranging from $m=5$ to $40$, we obtained values of $\rho_s$
differing only by $\lesssim 5\%$. By increasing $m$, indeed we observed a vary fast convergence 
to the asymptotic value of the stiffness.
This ensured us to obtain reliable results, even without pushing further the simulations 
to larger bond-link values.
On the other hand, one needs also to increase the truncation parameters 
with $m$, since too small values originate non-monotonic fluctuations
in the variational energy.~\cite{rossini11}
In particular, if an increase of $m$ is not accompanied 
by a gradual increasing of $p$ and $s$, the error bar in $\rho_s$ increases.

The scaling behavior of the spin-stiffness is analyzed in Fig.~\ref{fig:Stiff-scaling} for 
those values of $\Delta$ and $B$ for which there is a direct supersolid-to-solid transition. 
Data are shown for the lower critical field at $\Delta = 6$. 
Very close to the critical field $B_c$ the data are described accurately by a power-law 
behavior $\rho_s \sim (B-B_c)^{\beta_s}$. The value of the exponent is very sensitive to the location of the critical 
point, a change in its estimate on the third digit may change the value of the fitted exponent up to few percents. 
By fitting all the values up to the vertical bar in Fig.~\ref{fig:Stiff-scaling} 
and using a value of $B_c = 8.5052$, we get a best fit to the exponent of $\beta_s = 0.511$ which is
in very good agreement with the theoretical value $\beta_s = 0.5$ (dashed blue line).~\cite{note2}

\begin{figure}
  \begin{center}
    \includegraphics[width=8.5cm]{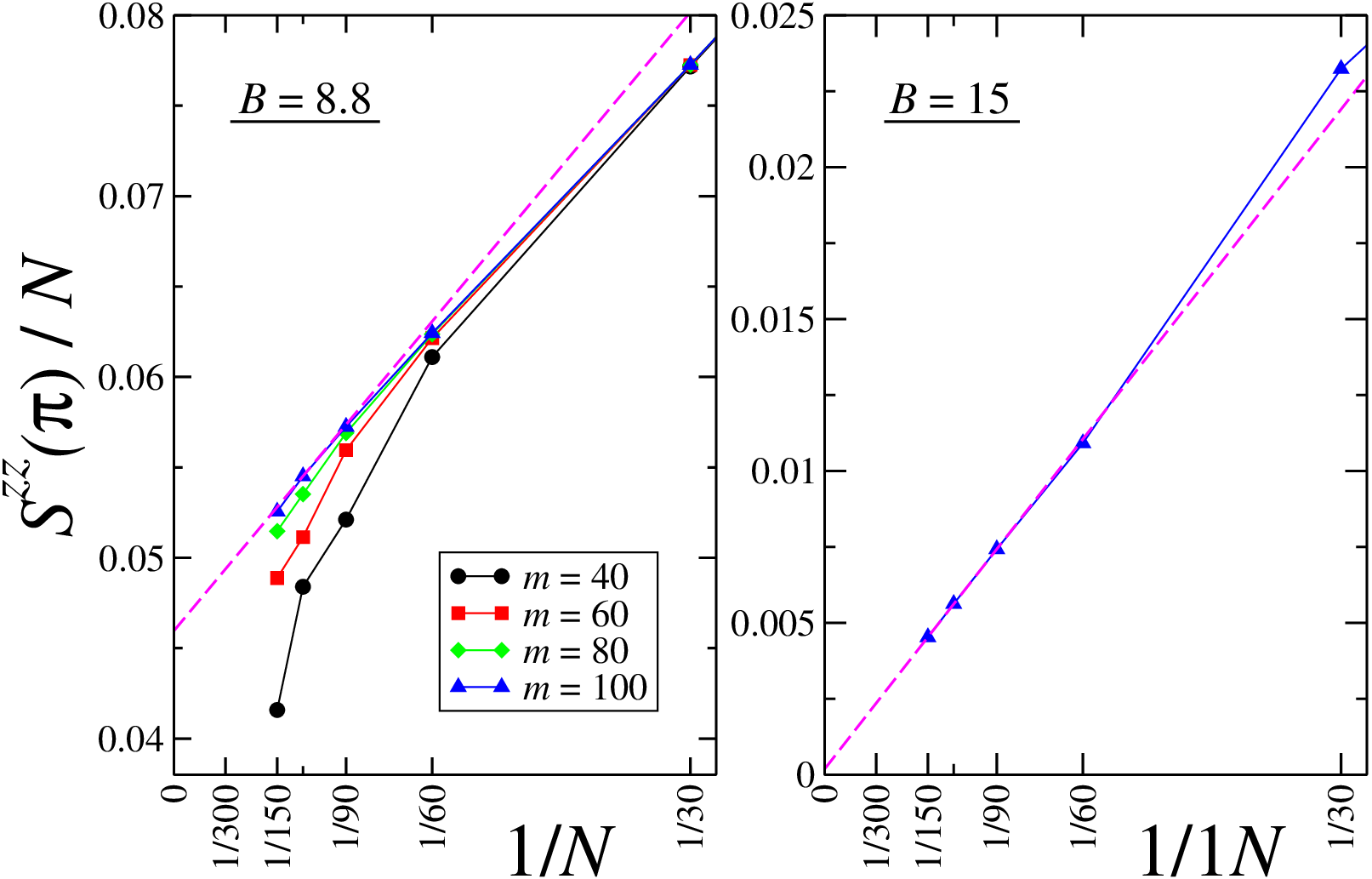}
    \caption{(color online). Spin structure factor $S^{zz}(\pi)$ as a function of the system size $N$ 
      for $\Delta = 6, \, D = \Delta/2$, while $B = 8.8$ (left panel) and $B=15$ (right panel);
      this has been obtained with MPS variational technique with OBC. Data are rescaled over $N$. 
      Dashed lines are linear fits of the three points at the largest sizes, for $m=100$.
      A finite value of ${\mathcal O}_{SDW} \approx 0.046 \pm 0.001$ can be obtained by extrapolating 
      the $N \to \infty$ value in the left panel. On the other hand, in the right panel 
      a value of ${\mathcal O}_{SDW} \approx 1.18 \times 10^{-4} \pm 10^{-4}$ 
      is extrapolated at the thermodynamic limit.
      This corresponds to a vanishing solid order parameter, within numerical accuracy
      given by the linear fits.}
    \label{fig:Solid_Spin}
  \end{center}
\end{figure}

\begin{figure}
  \begin{center}
    \includegraphics[width=\columnwidth]{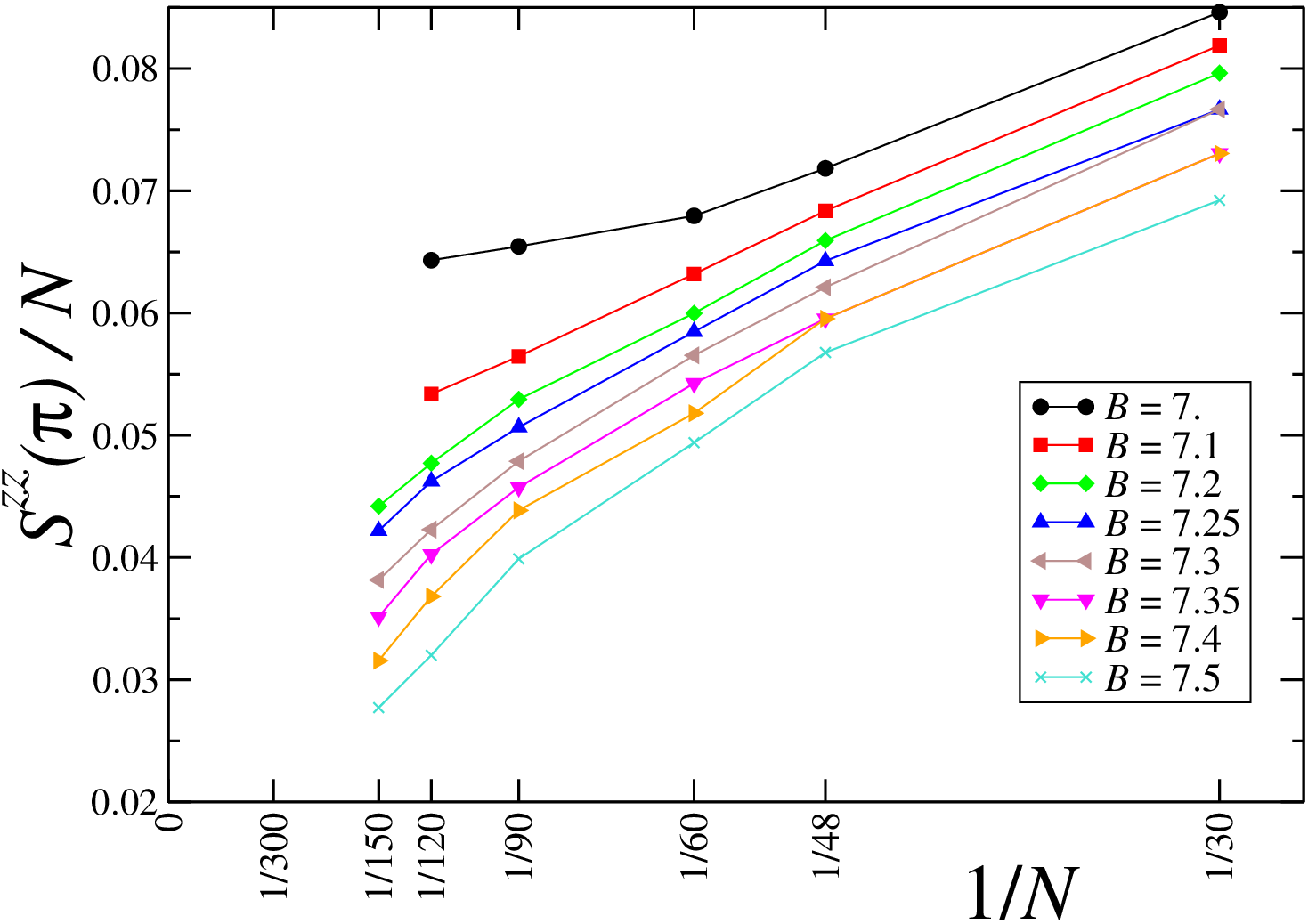}
    \caption{(color online). Spin structure factor $S^{zz}(\pi)$ as a function of the system size $N$ 
      at $\Delta = 5, \, D = \Delta/2$ and for different values of the external field. 
      At a value of the field $B_c \approx 7.35 \pm 0.075$ there is an upturn of the curves showing 
      that the system becomes solid.}
    \label{fig:Solid_Spin3}
  \end{center}
\end{figure}

The calculation of the solid order required much larger MPS matrix dimensions. 
However, as already mentioned, since for large systems boundary 
effects are negligible when detecting the solid order, we computed $S^{zz}(\pi)$ by resorting 
to a standard variational MPS algorithm with OBC, where much larger $m$ values are attainable.
To guarantee that our data are not qualitatively affected by boundary effects,
we compared $S^{zz}(\pi)$ of Eq.~\eqref{eq:StructureZZ}
with the one evaluated by summing up only over a fraction of the spins 
corresponding to the central part of the chain (say, $1/3$ of the total length). 
The location of the phase transition point, where the solid order parameter drops
from a finite to a vanishing value, do not change, even if the value of ${\cal O}_{SDW}$
inside the solid phase can be different. 

The results for the structure factor are reported in Fig.~\ref{fig:Solid_Spin} for two emblematic cases. 
The left panel is obtained by setting $B=8.8$ and $\Delta =6$: it corresponds to a
configuration which is well inside to the cusp of Fig.~\ref{fig:Stiff_Spin1} of the supersolid phase. 
For these values the system should hence exhibits a non-null solid order parameter ${\mathcal O}_{SDW}$:  
this is clearly evident in the left panel of Fig.~\ref{fig:Solid_Spin}, 
where the value ${\mathcal O}_{SDW}\approx 4.6 \times 10^{-2}$ is found by extrapolating 
numerical data for $N \to \infty$ from the linear behavior in $N$ of the quantity $S^{zz}(\pi)$. 
[Notice that the solid ordering can be extracted only for $m \sim 100$, since at low $m$ 
the data accuracy rapidly deteriorates for larger sizes]. 
On the other hand, the right panel of Fig.~\ref{fig:Solid_Spin} is obtained for $B=15$ and $\Delta =6$. 
It corresponds to a configuration which is far away from the supersolid region and for which 
the simulations of Ref.~\onlinecite{sengupta07b} predicted that 
no solid order should exist (indeed, the system is a superfluid there). 
This is confirmed by our simulations, where we observed $S^{zz}(\pi)/N \to 0$ in the thermodynamic limit, 
within numerical accuracy (Fig.~\ref{fig:Solid_Spin}, right panel). 

Finally we observe that, for values of the anisotropy $\Delta \lesssim 5.5$ 
in Fig.~\ref{fig:spin1_PhDiag}, there is a direct transition from the supersolid to the superfluid phase. 
In this case the transition is detected by the vanishing of the solid order parameter. 
In Fig.~\ref{fig:Solid_Spin3} we show the spin structure factor as a function of the system 
size for different values of the external field, fixing $\Delta = 5$. 
A scanning of this type for different values of $\Delta$ allows to complete the boundaries of the supersolid phase.

In conclusion, we analyzed the supersolid phase in a one-dimensional anisotropic spin-$1$ Heisenberg model 
in a transverse magnetic field, and single-site uniaxial anisotropy. By means of an MPS variational 
calculation with PBC, we showed how to determine the spin-stiffness and the 
structure factor, such to locate the supersolid in the phase diagram of the system and
find the critical exponent of the transition to the solid phase.
For our model of interest, the resulting portion of the phase diagram containing the supersolid phase 
is shown in Fig.~\ref{fig:spin1_PhDiag}.

We acknowledge very fruitful discussions with S. Peotta, P. Sengupta, and P. Silvi. This work was 
supported by the FIRB-IDEAS project, RBID08B3FM, EU Projects IP-SOLID and ITNNANO.



\begin{thebibliography}{99}

\bibitem{originalsupersolid}
  A.F. Andreev and I.M. Lifshitz, Sov. Phys. JETP {\bf 29}, 1107 (1969);
  A.J. Leggett, Phys. Rev. Lett. {\bf 25}, 1543 (1970);
  H. Matsuda and T. Tsuneto, Suppl. Prog. Theor. Phys. {\bf 46}, 411 (1970).

\bibitem{prokofiev07}
  N. Prokof'ev, Adv. Phys. {\bf 56}, 381 (2007).

\bibitem{kim04}
  E. Kim and M.H.N. Chan, Nature {\bf 427}, 225 (2004).

\bibitem{coldatoms}
  T. Lahaye {\it et al.}, 
  Rep. Prog. Phys. {\bf 72}, 126401 (2009);
  L. Pollet {\it et al.}, 
  Phys. Rev. Lett. {\bf 104}, 125302 (2010);
  F. Cinti {\it et al.}, 
  Phys. Rev. Lett. {\bf 105}, 135301 (2010).

\bibitem{ng06}
  K.-K. Ng and T. K. Lee, Phys. Rev. Lett. {\bf 97}, 127204 (2006).

\bibitem{laflorencie07}	
  N. Laflorencie and F. Mila, Phys. Rev. Lett {\bf 99}, 027202 (2007).

\bibitem{picon08}
  J.-D. Picon {\it et al.}, 
  Phys. Rev. B {\bf 78}, 184418 (2008).

\bibitem{sengupta07a}
  P. Sengupta and C.D. Batista, Phys. Rev. Lett. {\bf 98}, 227201 (2007).

\bibitem{sengupta07b}
  P. Sengupta and C.D. Batista, Phys. Rev. Lett. {\bf 99}, 217205 (2007).

\bibitem{peters}
  D. Peters, I.P. McCulloch, and W. Selke, Phys. Rev. B {\bf 79}, 132406 (2009);
  J. Phys: Conf. Ser. {\bf 200} 022046 (2010).

\bibitem{verstraete08}
  F. Verstraete, V. Murg, and J.I. Cirac, Adv. Phys. {\bf 57}, 143 (2008).

\bibitem{verstraete04}
  F. Verstraete, D. Porras, and J.I. Cirac, Phys. Rev. Lett. {\bf 93}, 227205 (2004).

\bibitem{sandvik07}
  A. W. Sandvik and G. Vidal, Phys. Rev. Lett. {\bf 99}, 220602 (2007).

\bibitem{pippan10}
  P. Pippan, S.R. White, and H.G. Evertz, Phys. Rev. B {\bf 81} 081103(R) (2010).

\bibitem{pirvu10}
  B. Pirvu, F. Verstraete, and G. Vidal, Phys. Rev. B {\bf 83} 125104 (2011).

\bibitem{rossini11}
  D. Rossini, V. Giovannetti, and R. Fazio, arXiv:1102.3562

\bibitem{fisher73}
  M.E. Fisher, M.N. Barber, and D. Jasnow, Phys. Rev. A {\bf 8}, 1111 (1973).

\bibitem{note1}
  The singular value decomposition of the product of a long chain of MPS transfer matrices 
  (a $m^2 \times m^2$ matrix) typically has singular values rapidly decaying to zero.~\cite{pippan10}
  For practical purposes, it is sufficient to take into account only $p \ll m^2$ of them, 
  without compromising the accuracy (neglected values contribute with terms 
  of the order of roundoff errors).
  In a similar fashion, the effective Hamiltonian on the MPS basis can be well
  approximated by expanding it via a singular value decomposition, keeping only 
  the contributions associated to its $s$ largest eigenvalues.

\bibitem{note2}
  For large values of $\Delta$ and for $D=\Delta/2$, the model in Eq.~\eqref{eq:Spin1}
  at low energies can be mapped onto an effective XX spin-$1/2$ chain in a transverse field,
  as shown in Ref.~\onlinecite{sengupta07b}.
  We analytically extrapolated the critical exponent $\tilde{\beta}_s$ for such effective model after 
  diagonalizing it in momentum space (in presence of a generic twist at the boundary).
  We found a theoretical value $\tilde{\beta}_s = 0.5$; this agrees with the numerically computed 
  value $\beta_s$, within our accuracy.

\end{thebibliography}
\end{document}